\documentclass[12pt]{article}
\usepackage{bbold}
\usepackage{graphicx}
\def \babar{{\sc BaBar}}
\def \bea{\begin{eqnarray}}
\def \beq{\begin{equation}}

\def \bra#1{\langle #1 |}
\def \eea{\end{eqnarray}}
\def \eeq{\end{equation}}

\begin{document}
\rightline{EFI 13-13}
\rightline{arXiv:1307.2550}
\bigskip
\centerline {{\bf HADRONIC AND RADIATIVE $D^*$ WIDTHS}
\footnote{Submitted to Phys.\ Rev.\ D.}}
\bigskip
 
\centerline{Jonathan L. Rosner~\footnote{rosner@hep.uchicago.edu.}}
\centerline{Enrico Fermi Institute and Department of Physics}
\centerline{University of Chicago, 5620 S. Ellis Avenue, Chicago, IL 60637}
\medskip
 
\begin{quote}
A recent measurement of the total $D^{*+}$ width, $\Gamma_{\rm tot}(D^{*+})
= (83.3 \pm 1.3 \pm 1.4)$ keV, is shown to be close to earlier predictions
based on the single-quark-transition hypothesis.  Those predictions are updated
using more recent masses and branching fractions to a value of 80.4 keV, with
a small uncertainty associated with the radiative branching fractions of
$D^{*0}$ and $D^{*+}$.  A prediction for the total width of $D_2^*(2460)$
and its partial width into $D^* \pi$ and $D \pi$ is also updated, and found
to be in agreement with experiment.
\end{quote}

\centerline{PACS numbers: 14.40.Lb, 13.25.Ft, 12.39.Hg, 12.39.Jh}
\bigskip

\centerline{\bf I.  INTRODUCTION}
\medskip

The \babar Collaboration \cite{Lees:2013uxa,Lees:2013zna} has recently
measured the natural line width of the $D^{*+}(2010)$ vector meson, obtaining
the very precise value $\Gamma_{\rm tot}(D^{*+}) = (83.3 \pm 1.3 \pm 1.4)$ keV.
The purpose of the present note is to update old predictions of this width
\cite{Eichten:1979ms,Rosner:1985dx,Rosner:1988jk}.  For other early predictions
see, e.g., Refs.\ \cite{Godfrey:1985xj} and \cite{Thews:1984tn}; a later
discussion may be found in Ref.\ \cite{Amundson:1992yp}.  The new value, 80.4
keV, based on recent inputs, is close to the observed one.  Predictions of
radiative $D^*$ partial widths and of decay properties of the tensor charmed
meson $D_2^*(2460)$ are also updated and found to be in agreement with
observation.

Predictions for hadronic decays of $D^*$ are treated in Sec.\ II, while those
of radiative $D^*$ decays are discussed in Sec.\ III.  These results are then
combined and compared with experiment in Sec. IV.  Predictions for decay
properties of $D_2^*(2460)$ are updated and compared with experiment in
Sec. V, while Sec.\ VI concludes.
\bigskip

\centerline{\bf II. HADRONIC $D^*$ DECAYS}
\medskip

The treatment of hadronic decays of the form $M^* \to M \pi$ follows the
single-quark-transition formalism introduced by Melosh \cite{Melosh:1974cu}
and applied by Gilman, Kugler, and Meshkov \cite{Gilman:1973hc,Gilman:1973kh}.
We relate the decays $D^* \to D \pi$ to the well-measured decays $K^*(890)
\to K \pi$ using kinematic factors from Ref.\ \cite{Eichten:1979ms}.  For
a meson composed of one light quark ($u$ or $d$) and one heavy quark
(treating $s$ and $c$ both as heavy), one expects
\beq
\Gamma(M^* \to M \pi) = \frac{p_\pi^3}{M^*} C^2 E_\pi E_M |A|^2~,
\eeq
where $E_\pi$ and $E_M$ are the energies of $\pi$ and $M$ in the $M^*$
center-of-mass system (c.m.s.), $C$ is an isospin Clebsch-Gordan coefficient,
and $A$ is an amplitude taken common to $K^* \to K \pi$ and $D^* \to D \pi$.

We take the properties of the decay $K^* \to K \pi$ from Ref.\ \cite{PDG}.
For the decay width, we ignore small $K^*$ radiative decay rates and take the
partial width into $K \pi$ to be the total width.  Averaging the values
\cite{PDG} $\Gamma(K^{*\pm}) = 50.8 \pm 0.9$ MeV and $\Gamma(K^{*0}) = 48.7
\pm 0.8$ MeV, we estimate $\Gamma(K^* \to K \pi) = 49.6 \pm 0.6$ MeV.
(A slightly higher value of $51.1 \pm 1.1$ MeV was taken in Ref.\
\cite{Rosner:1985dx}.)  Taking average pion, kaon, and $K^*$ masses, we
find $p_\pi = 289$ MeV, $E_\pi = 320$ MeV, $E_K = 574$ MeV, and hence
$|A|^2 = 1.00 \times 10^{-8}$ MeV$^{-3}$.

We now use $D$ and $D^*$ masses from Ref.\ \cite{PDG} (consistent within
errors with the more precise measurement \cite{Lees:2013uxa,Lees:2013zna} of
$M(D^{*+}) - M(D^0) = 145~425.8 \pm 0.5 \pm 1.8$ keV) to calculate the
corresponding quantities for $D^* \to D \pi$ decays.  We find $\Gamma(D^{*+}
\to D^0 \pi^+) = 54.8$ keV, $\Gamma(D^{*+} \to D^+ \pi^0) = 24.2$ keV, and
$\Gamma(D^{*0} \to D^0 \pi^0) = 34.7$ keV.  These values are compared with
ones obtained in Refs.\ \cite{Eichten:1979ms} and \cite{Rosner:1985dx} in
Table \ref{tab:dhad}.  Differences are due primarily to slightly different
values of $p_\pi$ and $\Gamma(K^* \to K \pi)$.


\begin{table}
\caption{Comparison of predictions for $D^* \to D \pi$ partial widths, in keV.
\label{tab:dhad}}
\begin{center}
\begin{tabular}{c c c c c c c} \hline \hline
Decay & \multicolumn{2}{c}{Ref.\ \cite{Eichten:1979ms}} & 
 \multicolumn{2}{c}{Ref.\ \cite{Rosner:1985dx}} & 
 \multicolumn{2}{c}{This work} \\
  & $p_\pi$ (MeV) & $\Gamma$ (KeV) & $p_\pi$ (MeV) & $\Gamma$ (KeV) 
  & $p_\pi$ (MeV) & $\Gamma$ (KeV) \\ \hline
$D^{*+} \to D^0 \pi^+$ & 38.9 & 53.4 & 39.2 & 56.9 & 39.3 & 54.8 \\
$D^{*+} \to D^+ \pi^0$ & 36.6 & 22.2 & 38.4 & 25.9 & 38.2 & 24.2 \\
$D^{*0} \to D^0 \pi^0$ & 45.3 & 43.4 & 44.1 & 39.7 & 42.9 & 34.7 \\
\hline \hline
\end{tabular}
\end{center}
\end{table}
\bigskip

\centerline{\bf III.  RADIATIVE $D^*$ DECAYS}
\medskip

The radiative decays $D^* \to D \gamma$ are among a large class of magnetic
dipole transitions between $^3S_1$ and $^1S_0$ states, all of which have been
successfully described within a nonrelativistic approach \cite{Rosner:1980bd,
Gasiorowicz:1981jz}.  Such M1 transitions from vector mesons $M^*$ to
pseudoscalar mesons $M$ composed of quarks $A \bar B$ proceed at a rate
\cite{Jackson:1976,Eichten:1978tg}
\beq \label{eqn:rad}
\Gamma(M^* \to M \gamma) = \frac{p_\gamma}{3 \pi} [\mu_A + \mu_B]^2 |I|^2~,
\eeq
where $p_\gamma$ is the photon momentum in the c.m.s., and $\mu_A$ and
$\mu_B$ are the magnetic moments of the quarks, $\mu_A = |e|Q_A/(2 m_A)$, with
$|e|Q_A$ denoting the charge of quark $A$.  The quantity $I = \bra{f} i
\rangle$ represents the overlap between initial and final wave functions.
$|I|^2$ is found to be about 1/2 for a large class of light-quark transitions
\cite{Rosner:1980bd,Gasiorowicz:1981jz} and for the charmed quark mass $m_c =
1662$ MeV assumed in Ref.\ \cite{Rosner:1985dx} gives rise to a prediction
$\Gamma(J/\psi \to \eta_c \gamma) = 2.32$ keV.  With the observed branching 
fraction ${\cal B}(J/\psi \to \eta_c \gamma) = (1.7 \pm 0.4)\%$
and observed total width $\Gamma_{\rm tot}(J/\psi) = (92.9 \pm 2.8)$ keV
\cite{PDG}, this gives an observed $\Gamma(J/\psi \to \eta_c \gamma) = (1.58
\pm 0.37)$ keV, or $|I|^2 = 0.68 \pm 0.16$.  A particularly well-determined
light-quark radiative decay width, $\Gamma(K^{*0} \to K^0 \gamma) = 116.5
\pm 9.9$ keV \cite{Carlsmith:1985}, when compared with the prediction of
Eq.\ (\ref{eqn:rad}), gives $|I|^2 = 0.52 \pm 0.04$ for the light-quark
masses \cite{Rosner:1980bd,Gasiorowicz:1981jz} $m_u = m_d = 310$ MeV/$c^2$,
$m_s = 485$ MeV/$c^2$.  The overlaps $|I|^2 < 1$ may be taken as a proxy for
relativistic effects ignored here.

Applying Eq.\ (\ref{eqn:rad}) to radiative $D^*$ decays, we find
\beq
\Gamma(D^{*+} \to D^+ \gamma) = (2.76~{\rm keV})|I|^2~,~~
\Gamma(D^{*0} \to D^0 \gamma) = (40.8~{\rm keV})|I|^2~.
\eeq
The suppression of the charged $D^*$ radiative decay width is due to the
partial cancellation of contributions of the charmed and $d$ quarks.  This
suppression was noted quite early \cite{Eichten:1979ms,Thews:1984tn}, and
was confirmed experimentally only some time later.
\bigskip

\centerline{\bf IV.  COMPARISON WITH EXPERIMENT}
\medskip

\begin{figure}
\begin{center}
\includegraphics[width=0.7\textwidth]{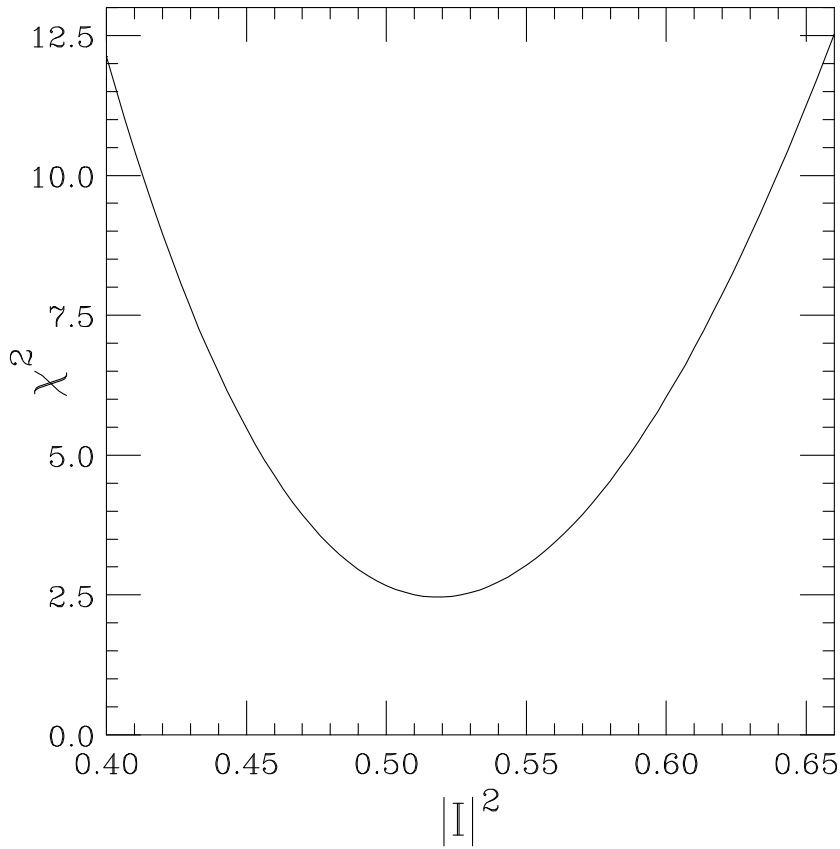}
\end{center}
\caption{Dependence of $\chi^2$ for fit to three $D^{*+}$ and two $D^{*0}$
branching fractions on the square $|I|^2$ of the overlap integral describing M1
transitions.
\label{fig:chsq}}
\end{figure}

We predict the following total widths for $D^{*+}$ and $D^{*0}$, parametrized
as functions of the square $|I|^2$ of the overlap integral describing M1
radiative transitions:
\beq
\Gamma_{\rm tot}(D^{*+}) = (79.0 + 2.76|I|^2){\rm~keV}~,~~
\Gamma_{\rm tot}(D^{*0}) = (34.7 + 40.8|I|^2){\rm~keV}~.
\eeq
Using the partial widths predicted in the previous two Sections, we may then
predict branching fractions as functions of $|I|^2$.  They may be compared
with the observed branching fractions \cite{PDG} and a $\chi^2$ formed based
on three $D^{*+}$ and two $D^{*0}$ branching fractions.  The result is shown
in Fig.\ \ref{fig:chsq}.  The minimum $\chi^2 = 2.46$ occurs for $|I|^2 =
0.52$, and is less than one unit above the minimum for a variation of $\pm
0.04$ about this value.

\begin{figure}
\begin{center}
\includegraphics[width=0.98\textwidth]{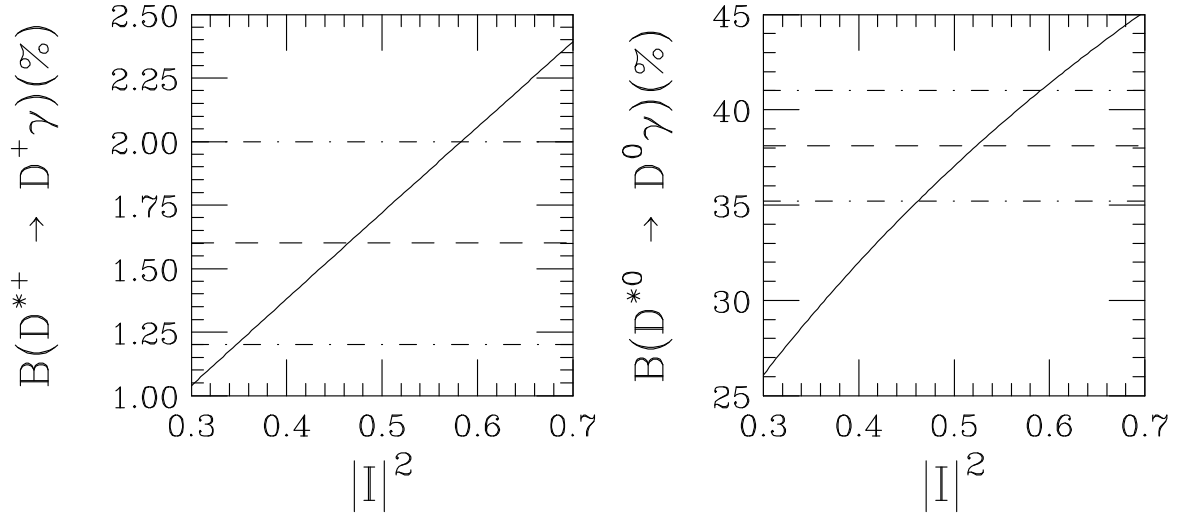}
\end{center}
\caption{Dependence of branching fractions for radiative $D^*$ decays on
square $|I|^2$ of overlap integral.  Horizontal dashed lines denote observed
central values of branching fractions \cite{PDG}, with $\pm 1 \sigma$ errors
denoted by dot-dashed lines.
\label{fig:brs}}
\end{figure}

The dependence of the branching fractions for the radiative $D^*$ decays
is shown in Fig.\ \ref{fig:brs}.  The major constraint on $|I|^2$ clearly
comes from ${\cal B}(D^{*0} \to D^0 \gamma)$

In Table \ref{tab:comp} we compare our predictions for $|I|^2 = 0.52 \pm 0.04$
with experiment.  The agreement is satisfactory.  The predicted $D^{*0}$
total width is far below an old upper bound \cite{Abachi:1988} of 2.1 MeV.
It would be an interesting experimental challenge to see if some very rare
$D^{*0}$ decay with a calculable decay rate could be used to provide
indirect information on the $D^{*0}$ total width.

\begin{table}
\caption{Comparison of predicted $D^*$ branching fractions and total widths
with experiment \cite{PDG}, for $|I|^2 = 0.52 \pm 0.04$.
\label{tab:comp}}
\begin{center}
\begin{tabular}{c c c} \hline \hline
Quantity & Prediction & Experiment \cite{PDG} (a) \\ \hline
${\cal B}(D^{*+} \to D^0 \pi^+)$ & $(68.1\pm0.1)\%$ & $(67.7 \pm 0.5)\%$ \\
${\cal B}(D^{*+} \to D^+ \pi^0)$ & $(30.1\pm0.1)\%$ & $(30.7 \pm 0.5)\%$ \\
${\cal B}(D^{*+} \to D^+\gamma)$ & $( 1.8\pm0.2)\%$ & $( 1.6 \pm 0.4)\%$ \\
$\Gamma_{\rm tot}(D^{*+})$ & $(80.5\pm0.1)$ keV & $(83.3\pm1.3\pm1.4)$ keV \\
${\cal B}(D^{*0} \to D^0 \pi^0)$ & $(62.0\pm1.7)\%$ & $(61.9 \pm 2.9)\%$ \\
${\cal B}(D^{*0} \to D^0\gamma)$ & $(38.0\pm1.7)\%$ & $(38.1 \pm 2.9)\%$ \\
$\Gamma_{\rm tot}(D^{*0})$ & $(55.9\pm1.6)$ keV & $< 2.1$ MeV \\
\hline \hline
\end{tabular}
\end{center}
\leftline{(a) Experimental value of $\Gamma_{\rm tot}(D^{*+})$ from
Ref.\ \cite{Lees:2013uxa,Lees:2013zna}.}
\end{table}
\bigskip

Ref.\ \cite{Amundson:1992yp} takes into account chiral loops and
calculates the total $D^{*+}$ width as a function of ${\cal B}(D^{*+} \to D^+
\gamma)$.  For any given value of this branching ratio, a range of $D^{*+}$
widths is obtained as a result of uncertainties in $D^{*0}$ branching fractions,
which amounted to 4.0\% at the time.  These have now been reduced to 2.9\% (see
Table \ref{tab:comp}).  At the time of Ref.\ \cite{Amundson:1992yp} only an
upper bound on ${\cal B}(D^{*+} \to D^+ \gamma)$ of 4.2\% was available.  With
the values in Table \ref{tab:comp} Ref.\ \cite{Amundson:1992yp} would now find
a range consistent with experiment:
\beq
64~{\rm keV} \le \Gamma_{\rm tot} \le 123~{\rm keV}~.
\eeq

\centerline{\bf V.  UPDATED PREDICTIONS FOR $D_2^*(2460)$}
\medskip

In Ref.\ \cite{Rosner:1985dx} we related D-wave decays of $K_2(1425)$ (now
called $K_2(1430)$) to the corresponding D-wave decays of the lowest-lying
spin-two charmed meson, assuming its mass was 2420 MeV/$c^2$.  This
corresponded to a $D^*\pi$ resonance whose spin and parity had not yet been
determined.  The state now called $D_2^*(2460)$ has a mass which, based on the
average of charged and neutral states in Ref.\ \cite{PDG}, we shall take to be
$2462.8 \pm 0.7$ MeV/$c^2$.  Scaling the prediction of \cite{Rosner:1985dx} to
the new $D_2$ mass, we then find
\beq
\Gamma(D^*_2 \to D \pi) = 27.4~{\rm MeV}~,~~
\Gamma(D^*_2 \to D^* \pi) = 19.4~{\rm MeV}~.
\eeq
Neglecting other possible decays (none are reported in Ref.\ \cite{PDG}),
we find a total width of 46.8 MeV, to be compared with the observed average
of the charged and neutral widths, $48.4 \pm 1.4$ MeV.  The predicted ratio of
$D \pi$ to $D^* \pi$ partial widths is 1.41, to be compared to the observed
value of $1.56 \pm 0.16$ \cite{PDG}.  Predictions in this range (under
varied assumptions) also have been obtained in Ref.\ \cite{Godfrey:1990}.

New data on the $D_2^*(2460)$ not included in the 2012 Particle Data Group
average have been reported by the ZEUS Collaboration at HERA \cite{ZEUS}:
\beq
\Gamma(D_2^{*0}) = (46.6 \pm 8.1^{+5.9}_{-3.8})~{\rm MeV}~,
\eeq
\beq
\frac{{\cal B}(D_2^{*0} \to D^+ \pi^-)}{{\cal B}(D_2^{*0} \to D^{*+} \pi^-)}
= 1.4 \pm 0.3 \pm 0.3~,
\eeq
\beq
\frac{{\cal B}(D_2^{*+} \to D \pi)}{{\cal B}(D_2^{*+} \to D^* \pi)}
= 1.1 \pm 0.4^{+0.3}_{-0.2}~.
\eeq
These results are also in accord with our predictions.
\bigskip
 
\centerline{\bf VI.  CONCLUSIONS}
\medskip

Experimental results continue to confirm some very early predictions of
charmed meson hadronic and radiative decays.  These predictions lie within the
scope of heavy quark effective theory but antedate it by a considerable
amount.  The refinement of Ref.\ \cite{Amundson:1992yp}, incorporating
contributions of chiral loops, also leads to predictions in accord with
experiment.
\bigskip

\centerline{\bf ACKNOWLEDGMENTS}
\medskip

I thank Uri Karshon for a helpful communication.  This work was supported in
part by the United States Department of Energy under Grant No.\ DE FG02
90ER40560.

\end{document}